      [1999/12/01 v1.4c Il Nuovo Cimento]
\documentclass{cimento}

\usepackage{graphicx}

\title{\textit{Swift} GRBs: the early afterglow spectral energy distribution}

\author{G.~Tagliaferri\from{ins:x}\ETC,
D.~Malesani\from{ins:y}\from{ins:yy},
S.D.~Vergani\from{ins:z}\from{ins:z2},
S.~Campana\from{ins:x},
G.~Chincarini\from{ins:x}\from{ins:xy},
S.~Covino\from{ins:x},
C.~Guidorzi\from{ins:x}\from{ins:xy},
A.~Moretti\from{ins:x},
P.~Romano\from{ins:x}\from{ins:xy},
L.A.~Antonelli\from{ins:x3},
M.~Capalbi\from{ins:x1},
M.L.~Conciatore\from{ins:x1},
G.~Cusumano\from{ins:x2},
P.~Giommi\from{ins:x1},
V.~La Parola\from{ins:x2},
V.~Mangano\from{ins:x2},
T.~Mineo\from{ins:x2},
M.~Perri\from{ins:x1}
\atque
E.~Troja\from{ins:x2}}

\instlist{
\inst{ins:x} INAF, Osservatorio Astronomico di Brera, via E. Bianchi 46, I-23807 Merate (Lc), Italy
\inst{ins:y} SISSA-ISAS, via Beirut 2-4, I-34014 Trieste, Italy 
\inst{ins:yy} Dark Cosmology Centre, Niels Bohr Inst., J. Maries vej 30, DK-2100 K\o{}benavn, Denmark
\inst{ins:z} Dunsink Observatory, Castleknock, Dublin 15, Ireland
\inst{ins:z2} School of Physical Sciences and NCPST, Dublin City University - Dublin 9, Ireland.
\inst{ins:xy} Universit\`a degli Studi di Milano-Bicocca, piazza delle Scienze 3, I-20126 Milano, Italy
\inst{ins:x3} INAF, Osservatorio Astronomico di Roma, via Frascati 33, I-00040 Monteporzio, Italy
\inst{ins:x1} ASI Science Data Center, via G. Galilei, I-00044 Frascati (Roma), Italy
\inst{ins:x2} INAF, IASF Palermo, via U. La Malfa 153, I-90146 Palermo, Italy
}
\PACSes{\PACSit{98.70.Rz}
\PACSit{98.70.Rz}{\ldots}}
\begin{document}

\maketitle

\begin{abstract}
We present the first results of a program to systematically study the optical-to-X-ray spectral
energy distribution (SED) of \textit{Swift} GRB afterglows with known redshift.
The goal is to study the properties of the GRB explosion and of the
intervening absorbing material. In this report we present the preliminary
analysis on 23 afterglows. Thanks to \textit{Swift}, we could build the SED at
early times after the GRB (minutes to hours). We derived the Hydrogen
column densities and the spectral slopes from the X-ray spectrum. We then
constrained the visual extinction by requiring that the combined optical/X-ray
SED is due to synchrotron, namely either a single power law or a broken power
law with a slope change by 0.5. We confirm a low dust-to-metal ratio, smaller
than in the SMC, even from the analysis of data taken significantly earlier
than previously possible. Our analysis does not support the existence of ``grey'' 
dust. We also find that the synchrotron spectrum works remarkably well to explain 
afterglow SEDs. We clearly see, however, that during the X-ray steep decay phases
and the flares, the X-ray radiation cannot be due only to afterglow emission.

\end{abstract}

\section{Introduction}

All evidences are that long-duration gamma-ray bursts (GRBs) are  associated
with the death of massive stars that explode as type-Ic supernovae. GRBs seem 
to be associated with the very energetic subclass of hypernovae, whose features
have been unambiguously identified in at least four cases: GRB\,980425,
GRB\,030329, GRB\,031203 and GRB\,060218
\cite{gal98,sta03,hjo03,mal04,cam06,pia06}. Therefore, long-duration GRBs
likely occur in the same regions where their massive progenitors were born and
rapidly evolve. If these regions are similar to the giant molecular clouds in
our Galaxy, then long-duration GRBs explode inside dense, dusty environments.
The huge energy emitted in the gamma-ray band during the prompt emission phase
is almost unaffected by absorption, allowing the detection of GRBs up to very
high redshifts (e.g. \cite{tag05,kaw06,jak06}). On the other hand, emission at
optical X-ray wavelengths is significantly affected by matter along the line of
sight (Hydrogen, gas, and dust). By studying afterglow spectra, we can then
infer the properties of the intervening matter, both in the proximity of the
explosion and along the line of sight. Furthermore, studying the afterglow
behaviour allows the investigation of the explosion physics. In fact both issues must
be handled together, since absorption modifies the observed spectrum, thus
affecting the comparison with models; on the other hand, the properties of the
intervening matter can be probed  effectively only with an estimate of the
intrinsic spectrum.

The standard model predicts that the afterglow radiation is produced by
synchrotron emission during the slowing down of a relativistic fireball which
impacts against  the surrounding material. This model has proven successful in
explaining the overall properties of afterglows, predicting, as observed,
power-law shapes for both the light curves and the spectra
\cite{MeszarosRees97}. Well-defined relations are set between the decay and
spectral power-law indices (the so-called closure relations), which have been
tested observationally. Despite an overall agreement, the wealth of accumulated
data has highlighted a complex situation, which has led many authors to
introduce several new ingredients, among which energy injection, radiative
losses, non-standard density profiles, angular structure, and varying
microphysical parameters \cite{Zhang06,Panaitescu06,Granot06}. The introduction
of these effects has been more or less capable to explain the new data, but has
partly reduced the predictive power of models, and the theoretical picture is
not yet fully estabilished. On the other hand, most of these solutions are
still based on the idea that the observed spectrum is due to synchrotron
emission. Independent of the details, the broad-band spectral shape can thus be
computed from robust first principles, and from optical through X-ray frequencies
it has the shape of a broken power law. With typical parameters, the break
frequency is interpreted as the cooling break. In this case, a robust
prediction is that the low- and high-energy spectral indices $\beta_1$ and
$\beta_2$ differ by exactly 0.5 ($F_\nu \propto \nu^{-\beta}$): $\beta_1 =
(p-1)/2$ and $\beta_2 = p/2$, where $p \approx 2$ is the index of the electron
energy distribution (e.g. \cite{sari98}). The position of the break frequency can 
however vary significantly from burst to burst (and it evolves with time for each case).

ISM intervening material affects the observed GRB spectrum by selectively extinguishing
the radiation. At X-ray frequencies, the absorption is due to metals (either in
the gas or dust phase), which affect mostly the low-energy range (below
$\approx 0.5$~keV). At optical and ultraviolet wavelengths, the spectral shape
is modified by dust. GRBs are very interesting sources for probing the
interstellar medium in the Universe, for a number of reasons. First, they are
very bright. Second, the intrinsic spectral shape is simple enough to allow a
reliable determination of the absorption amount. They are also observable
across a wide range of wavelengths allowing to probe complementary aspects of
the ISM. A lot of work has been already carried out in this respect. Optical
and X-ray spectra have shown that GRBs explode in dense environments, as it is
expected for sources located inside star-forming regions
\cite{Vreeswijk04,campa05}. The metallicity is generally low ($Z \sim
0.1Z_\odot$), as inferred from both absorption-line measurements
\cite{Vreeswijk04,Berger06} and from integrated spectra of a few host galaxies
\cite{Prochaska04,Klaas06}. Furthermore, there is very little dust content
(e.g. \cite{Fynbo01}), actually much less than expected even for such low
metallicities \cite{GalamaWijers01}.

To gather information on the physical properties of the afterglows and of the
intervening absorbing material, we started a program to study the
optical-to-X-ray spectral energy distribution (SED) of \textit{Swift} GRBs 
with known redshift. With respect to previous works (see e.g. 
\cite{Stratta04,nardi06,Kann06,Starling06}), thanks to \textit{Swift} we can
now extend the study to much earlier epochs. We can therefore check if there is
evolution in the properties of both the absorbing material and the afterglow
emission.

\section{Absorption and SEDs}

The dust content (parametrized by the optical extinction $A_V$) is usually
determined by fitting the photometric spectrum with an absorbed power law. This
procedure, however, is strongly sensitive to the adopted extinction law, which
is poorly known for the high-redshift ISM surrounding GRB sources. The most
remarkable feature of GRB dust is the lack of the 2150~\AA{} bump ubiquitously
observed along Milky Way sight lines, leading to a preference for an SMC-like
extinction curve \cite{Stratta04,Kann06}. This makes it even more difficult to
estimate the amount of total absorption, since a featureless extinction curve
can be hardly disentangled from a power-law shape unless good-quality
photometry is available.

\begin{figure} \begin{minipage}{0.5\textwidth}\scriptsize
\begin{tabular}{l@{}l|l@{}l|l@{}l}\hline GRB     &$z$    &    GRB     &$z$   
&    GRB     &$z$    \\ \hline \bf 050126  &1.29   &    050814  &5.3    &\bf
060210  &3.91   \\ 050223  &0.59   &\bf 050820A &2.612  &    060218  &0.03   \\
\bf 050315  &1.949  &    050826  &0.297  &\bf 060223A &4.41   \\ \bf 050318 
&1.44   &\bf 050824  &0.83   &\bf 060418  &1.49   \\ \bf 050319  &3.24   &\bf
050904  &6.295  &    060502  &1.51   \\ \bf 050401  &2.90   &\bf 050908 
&3.344  &    060505  &0.089  \\ \bf 050408  &1.236  &\bf 050922C &2.199  &   
060510B &4.9    \\ \bf 050416A &0.653  &    051016B &0.936  &    060512 
&0.443  \\ \bf 050505  &4.27   &    051109A &2.346  &    060522  &5.11   \\ \bf
050525A &0.606  &    051109B &0.08   &    050526  &3.21   \\ \bf 050603 
&2.821  &\bf 051111  &1.55   &    060604  &2.68   \\ \bf 050730  &3.967  &   
060115  &3.53   &    060605  &3.7    \\ 050802  &1.71   &\bf 060124  &2.297 
&    060607A &3.082  \\ \bf 050803  &0.422  &    060206  &4.048  &\bf 060614 
&0.125  \\ \hline \end{tabular} \end{minipage}\hfill%
\begin{minipage}{0.45\columnwidth}
\includegraphics[width=\columnwidth,origin=lb]{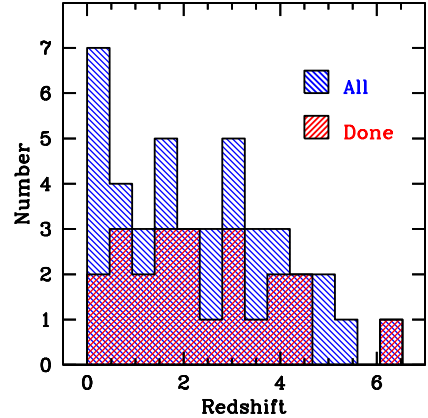} \end{minipage}
\caption{Left: list of GRBs included in our sample; bursts in boldface font
were analyzed in this preliminary work. Right: redshift distribution of the
full and analyzed samples. The average redshift is comparable for the two
distributions ($\langle z \rangle = 2.32$ and 2.34).\label{fg:z_distrib}}
\end{figure}

We present here a complementary approach to determine the dust content along
GRB sight lines, by modelling the combined optical and X-ray spectral energy
distribution, under the basic assumption that it is described by a synchrotron
spectrum. This task has been already performed by several authors using a
number of afterglows in the pre-\textit{Swift} era \cite{Stratta04,Starling06}.
Our aim is to carry on a full study of the \textit{Swift} sample. This kind of
analysis is potentially suffering from a selection bias, namely the requirement
that the burst is detected at optical wavelengths. Dusty afterglows more
frequently escape detection and, even if detected, they may lack detailed
photometric and spectroscopic studies. \textit{Swift} provides rapid and
precise GRB triggers, allowing a much more efficient follow up, therefore
effectively reducing this bias.

\begin{figure}
\includegraphics[width=0.5\columnwidth]{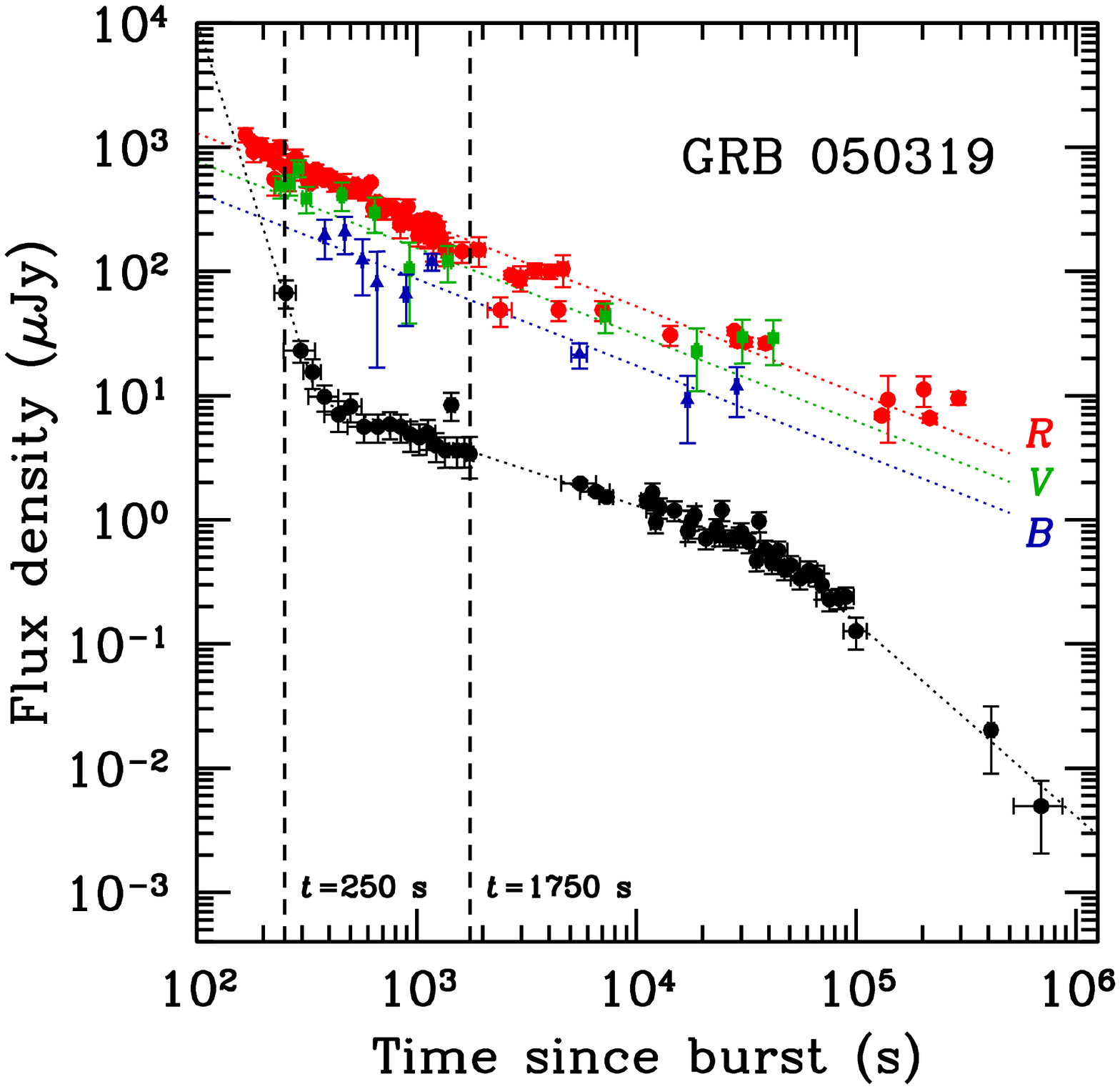}%
\includegraphics[width=0.5\columnwidth]{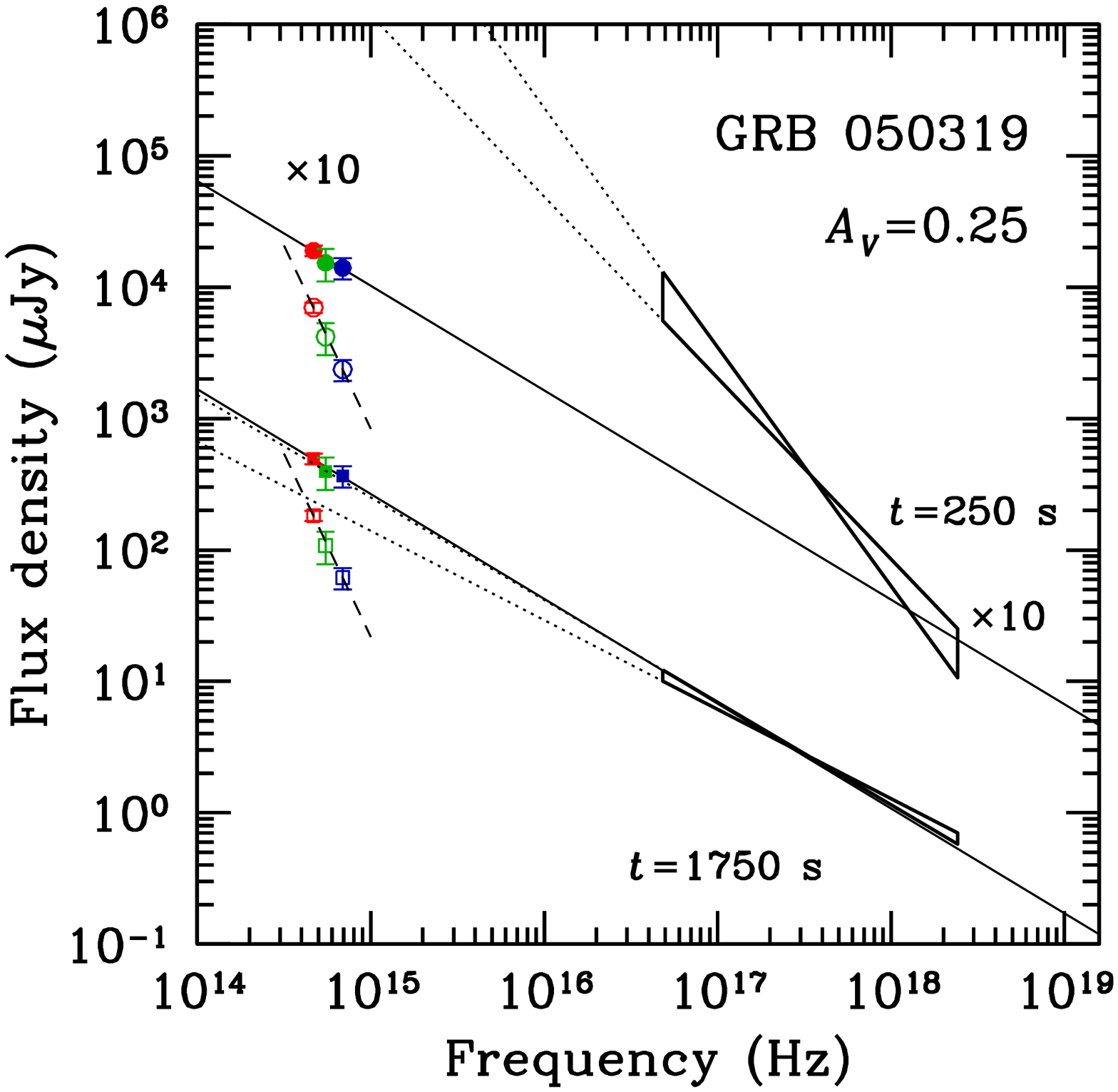}%
\caption{Left: optical and X-ray light curves of GRB\,050319 (data from
\cite{Cusumano06,Wozniak05,Mason06,Quimby06}). The vertical dashed lines mark
the epochs at which we computed the SED (observer frame). Right: SED of
GRB\,050319 at the two epochs marked in the left panel. Empty symbols are not
corrected for the host extinction, while filled symbols are. The solid and
dotted lines indicate the extrapolation of the optical and X-ray spectra,
respectively.\label{fg:050319}}
\end{figure}

To build the optical-to-X-ray SED we interpolated at a common time (from a few
minutes to hours after the GRB) optical and X-ray data. We took optical data
from the literature (including the GCNs) and from our own VLT and TNG MISTICI
data, after rejecting clearly discrepant or miscalibrated points. We decided 
not to extract SEDs during complex phases of the light curves, rather we chose
appropriate epochs with the best available spectral coverage. To avoid
degeneracy in the modeling of the dust properties, we also selected bursts with
measured (spectroscopic) redshift.

The XRT data were analysed following standard procedures.
We concentrated mainly on the X-ray data obtained during the first two \textit{Swift} 
orbits after the burst, from $\sim 100$~s up to 2--3~hr (although we looked at the 
full light curve and in some cases extracted also SEDs at later times).
In these early phases, the X-ray light curves are
characterised by a complex behaviour, following the usual steep-flat-steep
sequence, often with superimposed flares (e.g. \cite{chinca06,nouse06,chinca07};
Fig.~\ref{fg:050319}--\ref{fg:050416A}). When relevant, we accumulated different spectra for each
of these phases. Then we fitted them assuming an absorbed power-law model,
considering the absorption both in the Milky Way and local to the GRB.  This model
always provided a good fit to our data. 

To measure the optical extinction, we started by computing the X-ray spectral
slope $\beta_{\rm X}$ through fits to the XRT spectra. We then computed the
dust extinction by requiring either that {\it a)} the optical and X-ray spectra
lie on the same power-law component ($\beta_{\rm opt} = \beta_{\rm X}$) or that
{\it b)} the cooling frequency lies between the two bands, so that $\beta_{\rm
opt} = \beta_{\rm X} - 0.5$. To choose among the two possibilities, we compared
the relative normalizations of the optical and X-ray components: for example,
in case {\it a)} it is also necessary that the optical flux matches the
extrapolation of the X-ray spectrum. If both solutions do not work, this would
imply either a more complex spectral shape or a peculiar extinction curve. In
our sample, however, all cases but one fit the simplest model. Note that we did
{\it not} fit the optical data to find the extinction, but imposed a correction
to match the constraint from the X-ray slope. To model the extinction, we adopted 
both the SMC extinction curve \cite{Pei92} and the ``attenuation curve'' derived 
for starburst (SB) galaxies by \cite{Calzetti00}. Our method has the advantage of
being less sensitive to small errors in the photometric data, since the overall
extinction is constrained by the knowledge of the intrinsic slope.

\section{Results} 
 
The sample we have selected contains 42 \textit{Swift} long-duration GRBs with
spectroscopic redshift up to 2006 June. In this report we present our
preliminary results on 23 of them. Fig.~\ref{fg:z_distrib} shows the redshift
distribution of the full sample and of the analyzed one. 

\subsection{The intervening matter}

\begin{figure}
\includegraphics[width=0.5\columnwidth]{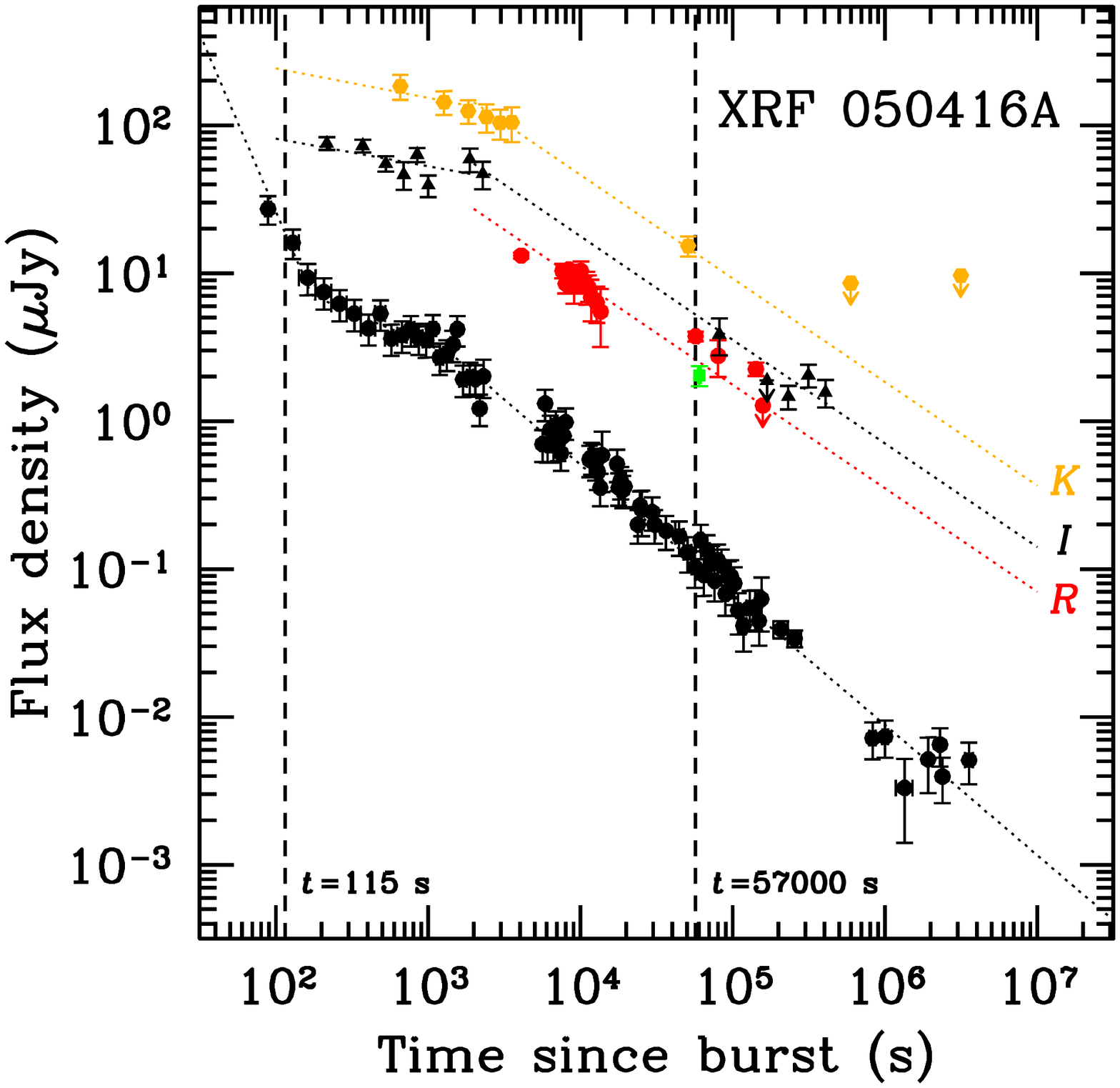}%
\includegraphics[width=0.5\columnwidth]{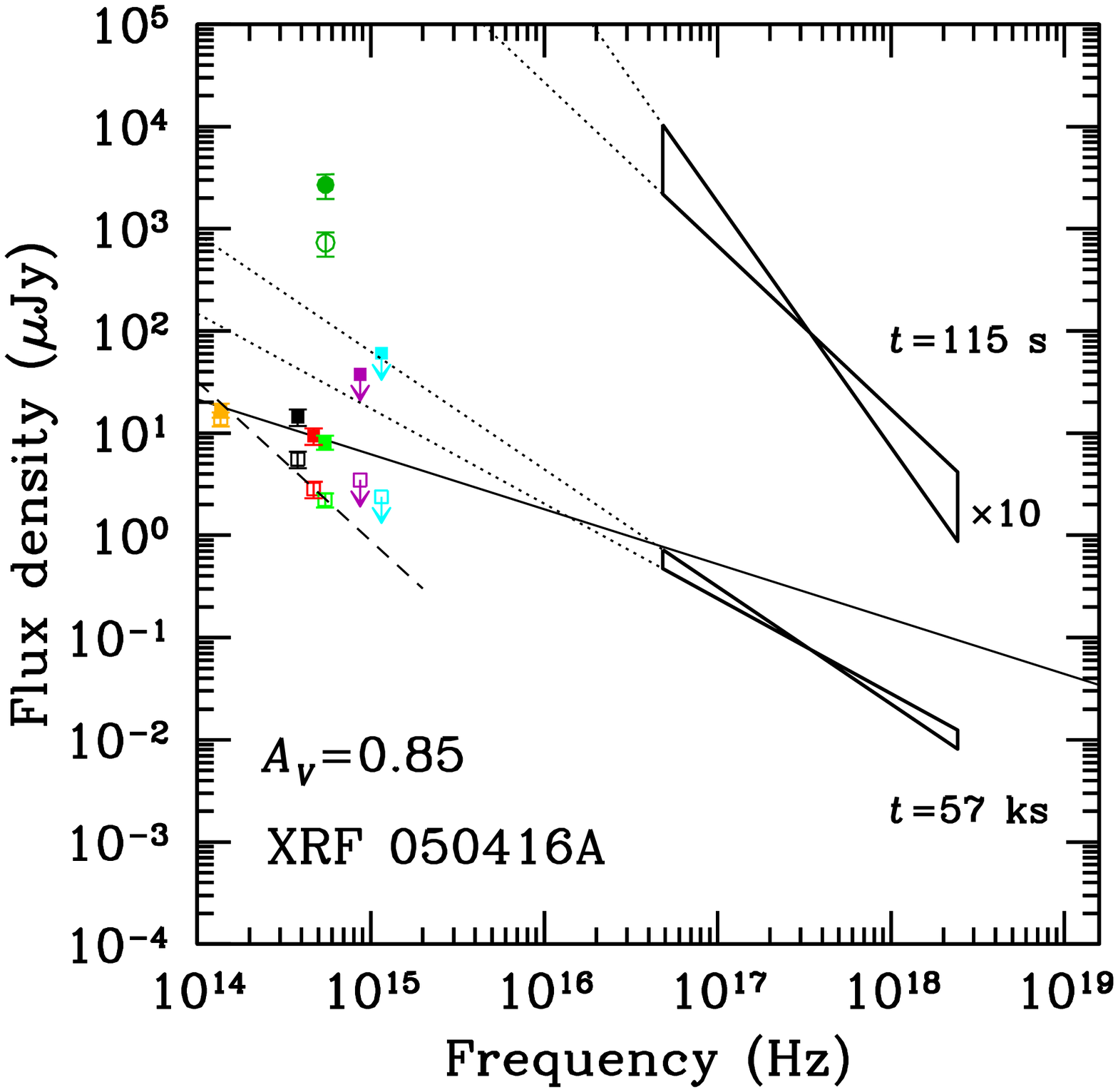}%
\caption{Left: optical and X-ray light curves of XRF\,050416A (data from
\cite{Mangano07,Holland07,Soderberg06}). The style convention are the same as
in Fig.~\ref{fg:050319}.\label{fg:050416A}}
\end{figure}

As an example, we show in Figs~\ref{fg:050319} and \ref{fg:050416A} the light
curves and SEDs of GRB\,050319 and XRF\,050416A. In the X-ray band, GRB\,050319
shows the canonical steep-flat-steep behaviour (Fig.~\ref{fg:050319}, left panel;
\cite{Cusumano06}).
At optical wavelengths the behaviour is described by a single power law.
Focussing on the late SED (taken during the flat X-ray phase), the presence of
dust is evidenced by the steep observed spectral index $\beta_{\rm opt} = 2.8$
(dashed line, right panel). However, after setting $A_V = 0.25$~mag, the
optical spectrum is much less steep and lie on the same power-law of the X-ray
spectrum. This interpretation, i.e. that the two component are due to the same
power law, is confirmed by the consistency of the temporal decay slopes in
the two bands (for $t < 30$~ks). Fig.~\ref{fg:050416A} shows the light curve
and SED of XRF\,050416A. Again, focussing on the later SED, the dust effect is
evident, but the cooling frequency was in this case lying between the optical
and X-ray bands, at $\nu \sim 2 \times 10^{16}$~Hz.

\begin{figure}
\includegraphics[width=0.5\columnwidth]{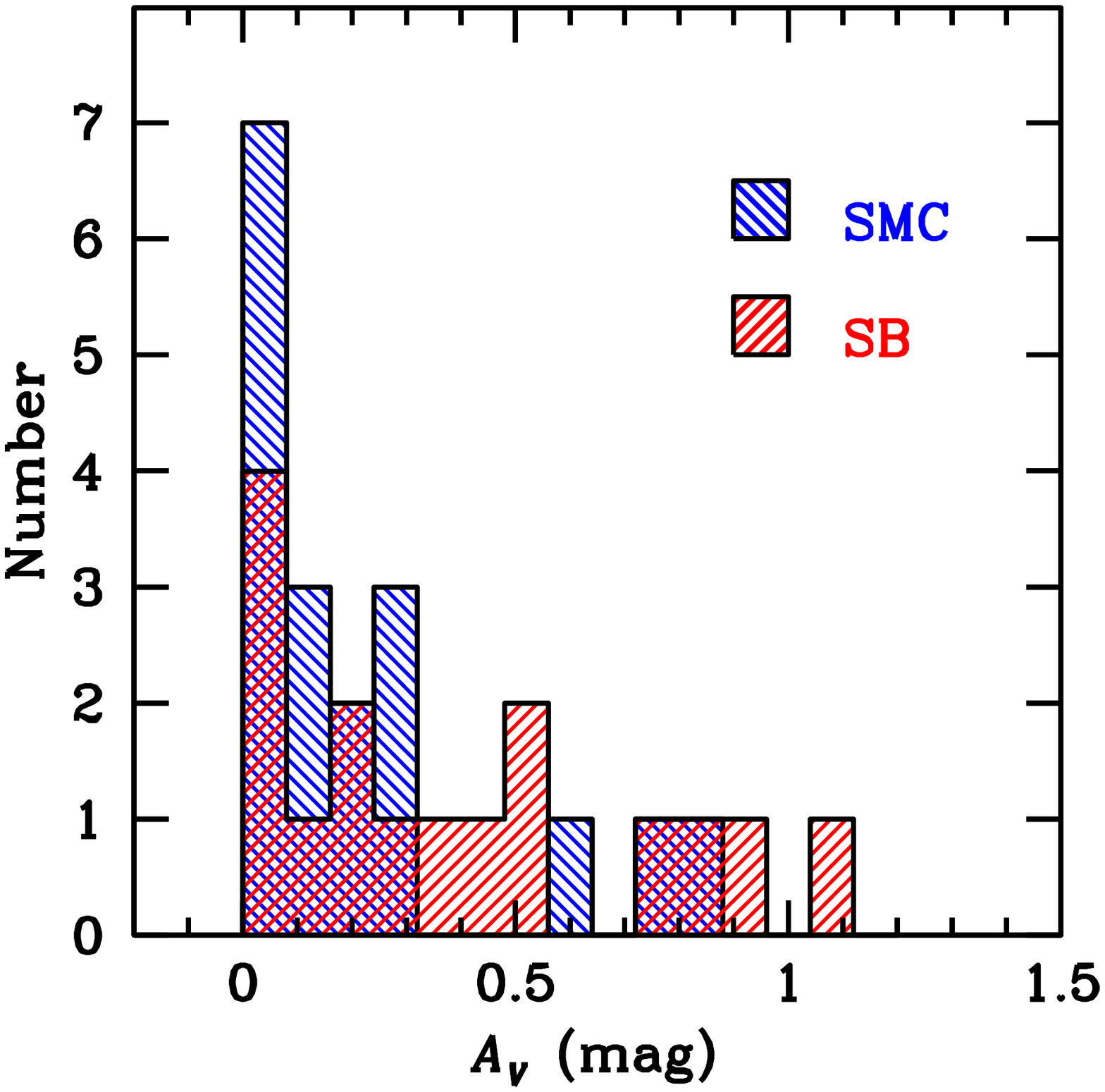}%
\includegraphics[width=0.5\columnwidth]{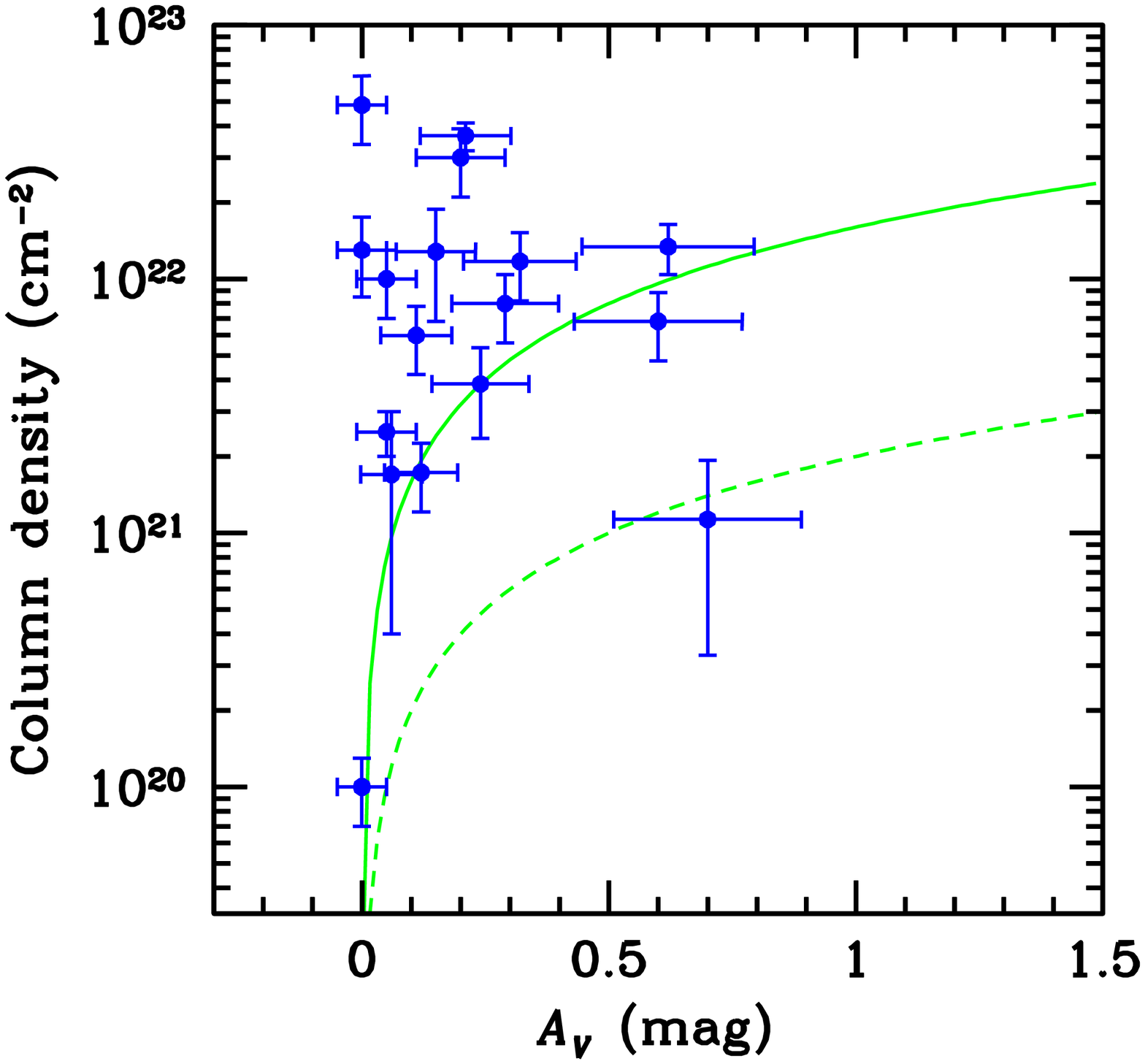}
\caption{Left: distribution of the rest-frame $A_V$ measured for our sample,
assuming an SMC or starburst (SB) extinction curve. Right: comparison of the
rest-frame $A_V$ and $N_{\rm H}$ (assuming an SMC extinction curve). The solid
line shows the SMC dust-to-gas ratio, and the dashed line shows the same ratio
when normalized to Solar metallicity.\label{fg:AV}}
\end{figure}

Figure~\ref{fg:AV} (left panel) shows the distribution of the measured $A_V$
with the two adopted extinction curves. The average values are $\langle A_V
\rangle = 0.22 \pm 0.26$ and $0.39 \pm 0.33$~mag for the SMC and SB (Calzetti)
curves, respectively. The SB extinction curve is flatter, so that on the
average a larger $A_V$ is needed in order to obtain the same reddening. In a
few cases, however, we could find no solution adopting the SB curve. Our
analysis, therefore, does not support the existence of ``grey'' dust. In
particular, our data never require a flat law $A(\lambda) \sim \mathrm{const}$
\cite{Maiolino01}. The average $A_V$ is consistent with pre-\textit{Swift}
values \cite{Kann06}, indicating that also with the extended \textit{Swift}
sample we have not yet probed the space of extinguished afterglows, although
sampling much earlier phases than before. We caution, however, that the bursts
in our sample all have a measured redshift, so they could still suffer from
some selection effect against faint (and more likely extinguished) afterglows
(the analysis of the few dark bursts with redshift is underway).

The right panel of Fig.~\ref{fg:AV} shows a comparison between the rest-frame
$A_V$ (using the SMC extinction curve) and the Hydrogen column densities
$N_{\rm H}$ as measured from the X-ray spectra (assuming Solar abundances). The
(logarithmic) average is $N_{\rm H}/A_V = 3 \times 10^{22}$
cm$^{-2}$~mag$^{-1}$, with a scatter of 0.55 dex. The solid line represents the
dust-to-gas ratio as measured in the SMC. Taken at face value, the column
densities are on the average larger than expected from this relation
\cite{GalamaWijers01,Stratta04}. We note also that our $N_{\rm H}$ values have
been computed assuming Solar metallicity, while the SMC has a lower value
($\approx Z_\odot / 8$). The dashed line shows the relation expected after
normalizing the dust-to-gas ratio to Solar metallicity, further exacerbating
the discrepance. Thus, the ISM medium in GRB host galaxies is different than
that of the SMC. Several explanations have been proposed to explain the low
dust content along GRB sight lines, including dust destruction
\cite{gala01,WaxmanDraine00,PernaLazzati02} due to the intense UV flux (either
from the GRB or from the neighbouring hot stars), or the young age of the
stellar populations (no time enough for dust formation)
\cite{Watson06,Campana06}. Alternatively, it is possible that the dust optical
properties are significantly different from the local templates. Some
suggestions in this direction have come from the analysis of optical absorption
lines \cite{Savaglio04}, even if our study does not favor flat extinction
curves. It could also be that these dust grains are smaller.

\subsection{GRB physics}
 
\begin{figure}
\includegraphics[width=0.33\columnwidth]{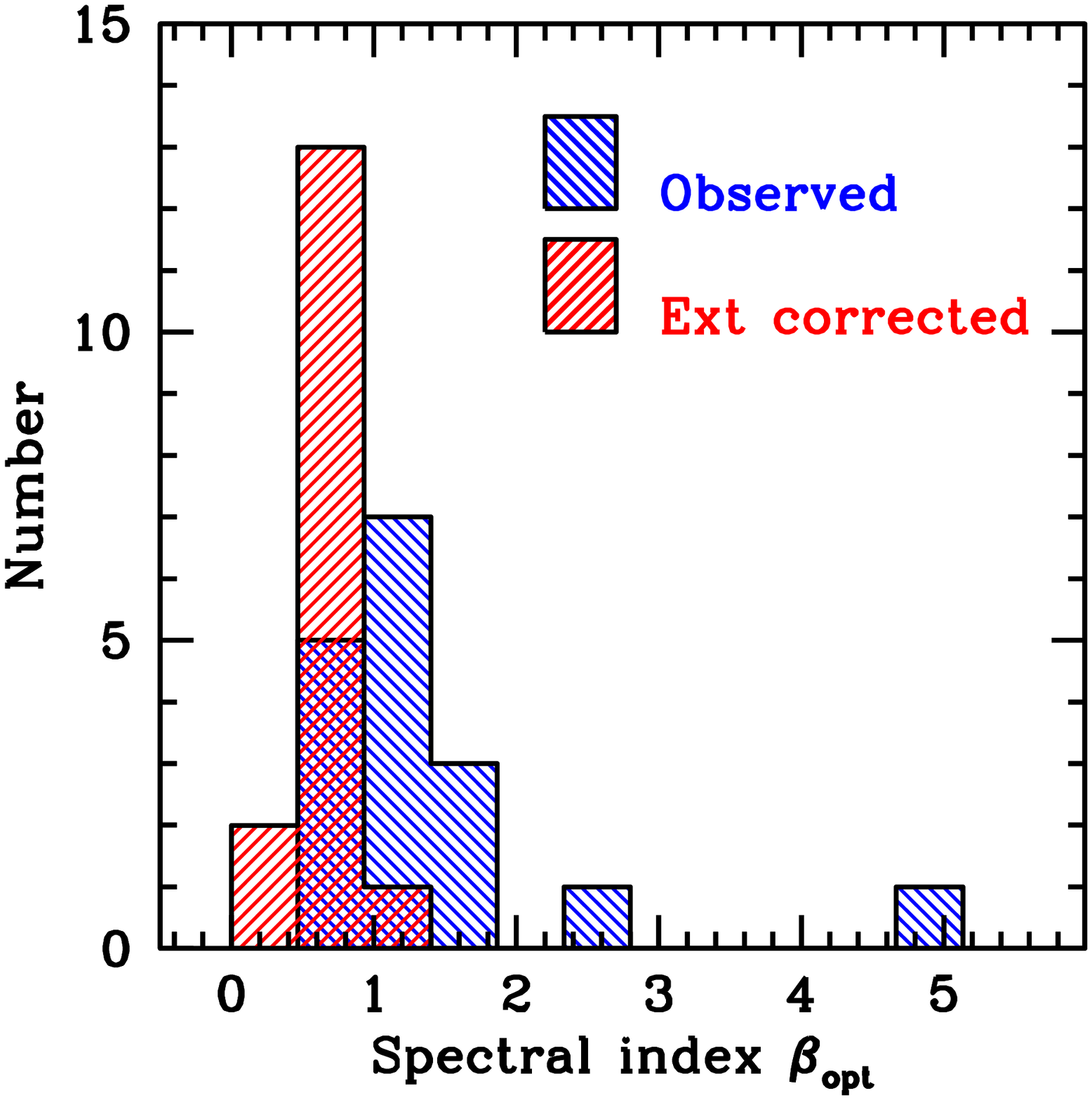}%
\includegraphics[width=0.33\columnwidth]{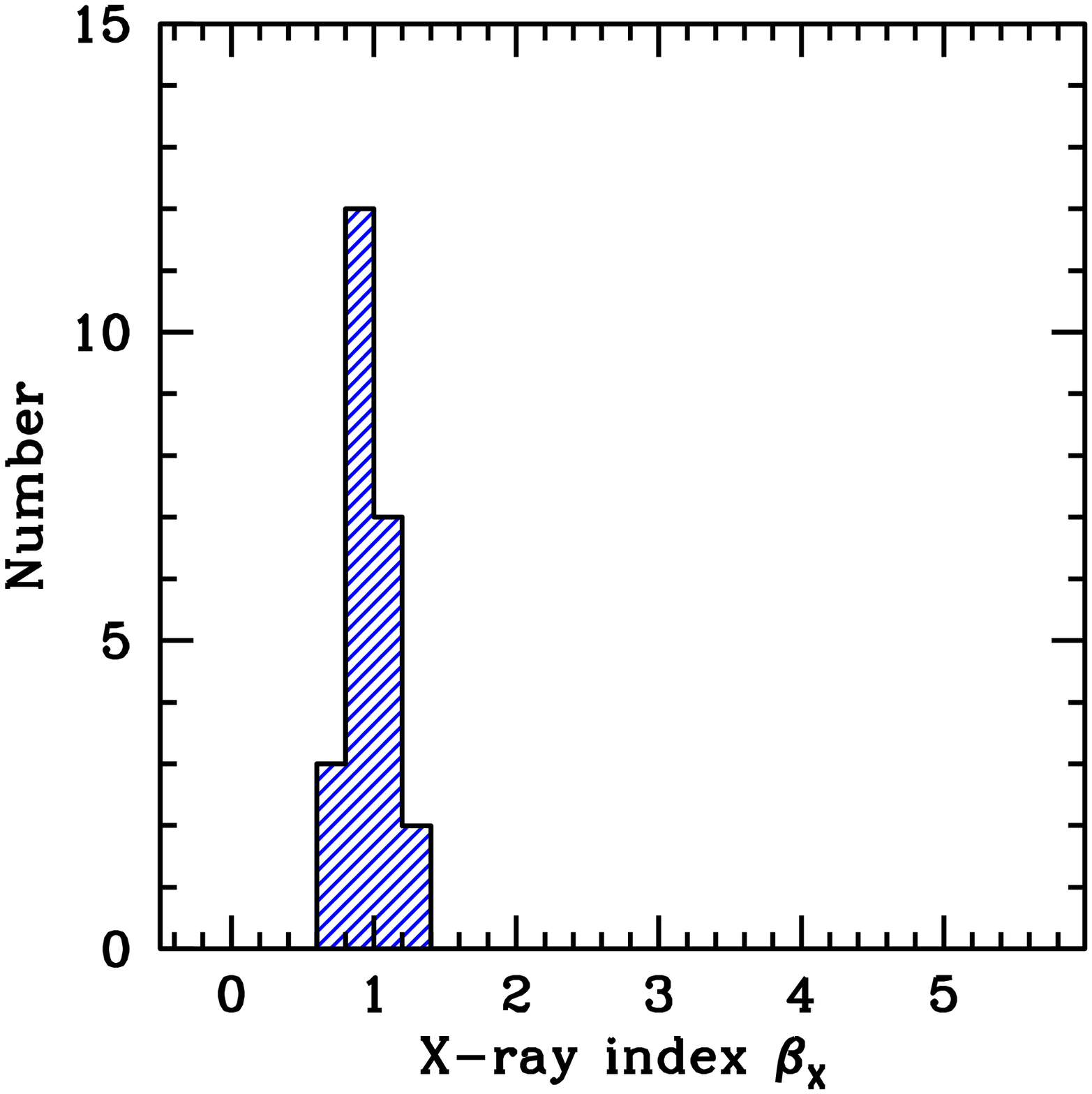}%
\includegraphics[width=0.33\columnwidth]{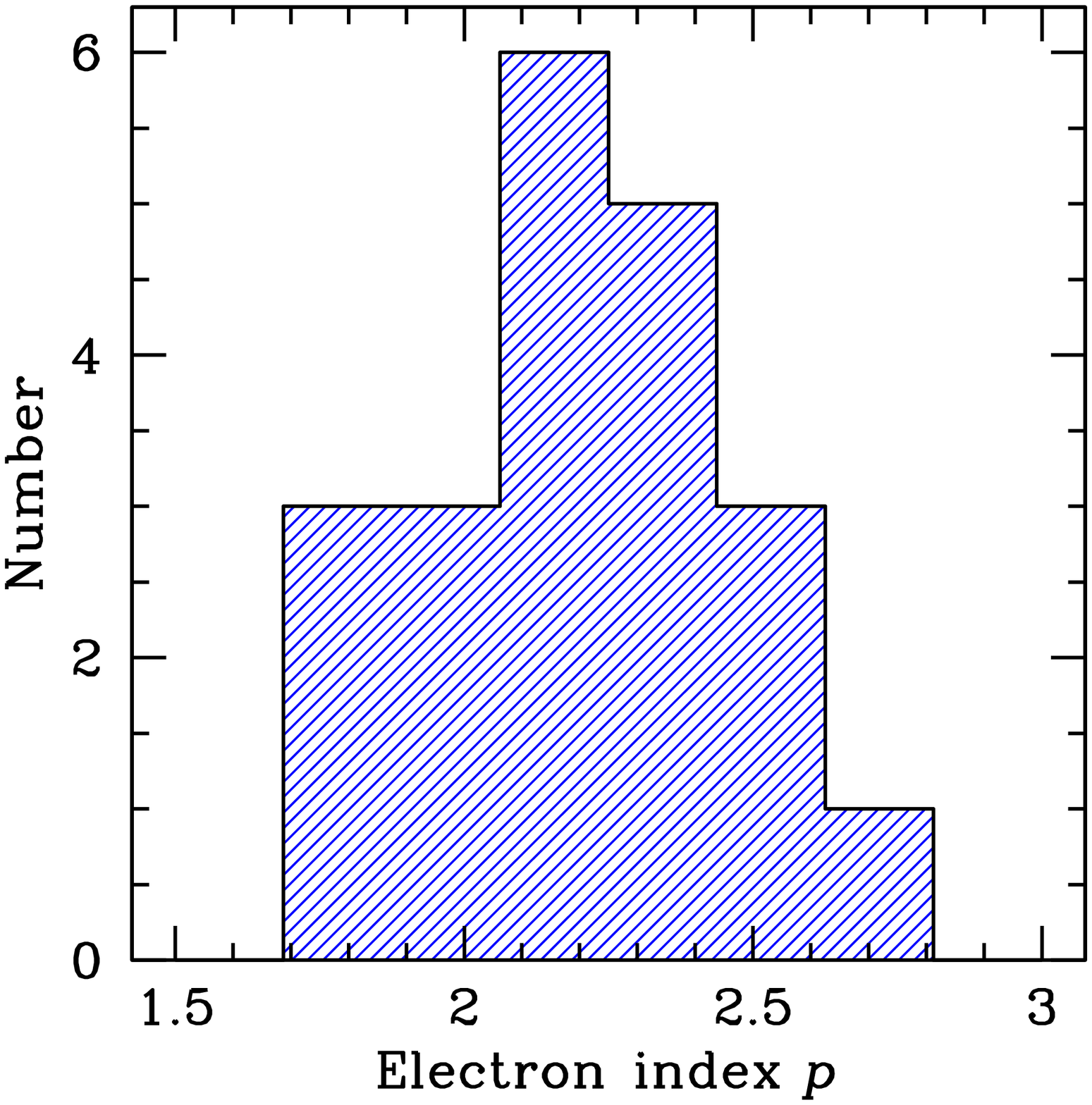}
\caption{Left and middle panels: distributions of the optical and X-ray
spectral indices. Right panel: distribution of the electron power-law
index.\label{fg:physics}}
\end{figure}

The first, interesting result is that the synchrotron spectrum fits remarkably
well. In the analyzed sample, only one case (GRB\,050904) could not be modeled
with a single synchrotron spectrum. In particular, there is no need to
introduce an inverse Compton component (which is anyway expected to contribute
only at later times, see \cite{harri01}). Thus, in contrast to the complexity in the temporal
domain, a simple spectral model works well for most afterglows.
Figure~\ref{fg:physics} shows the distribution of the spectral indices in the
optical and X-ray ranges, as well as the inferred electron distribution power law
index $p$. The observed optical slopes span a broad range, but after dust
correction they cluster around the typical value $\beta_{\rm opt} \approx 0.7$,
with a dispersion similar to that in the X rays. The electron index $p$
(computed as $p = 2\beta$ or $p = 2\beta + 1$ according to the location of the
cooling frequency) is also clustered around its ``canonical'' value: $\langle p
\rangle = 2.2 \pm 0.3$. Only $4/22$ bursts have $p < 2$.

With the temporal and spectral decay indices, we could then check the closure
relations. Of the 23 analyzed cases, 6 are fitted by an ISM model (and not by a
wind), while 3 are fitted by a wind (and not by a ISM); for 13 the available
data are consistent with both solutions, while for 1 no case works. For 5
bursts energy injection (or any other suitable mechanism) was required to
explain a flat phase in the X-ray light curve. For a few bursts, the SED shows
the cooling frequency to be close to or inside the XRT range ($\nu_{\rm c}
\approx 0.1$--0.5~keV). This suggests the possibility to test afterglow models
by measuring the time dependence (if any) of $\nu_{\rm c}$ (see also
\cite{ButlerKocevski06}).

There is another interesting feature evidenced by our SEDs. Both during the
X-ray flares and during the inital steep decay phases, the optical-to-X-ray
spectral index $\beta_{\rm OX}$ can be extremely hard (see the early SEDs of
GRB\,050319 and XRF\,050416A in Figs.~\ref{fg:050319} and \ref{fg:050416A}).
This is a confirmation also from the spectral point of view that the X-ray
radiation is not (or not only) afterglow emission during these phases, but is
likely of internal origin, as already suggested on the basis of temporal
properties (for a recent discussion on this subject see also \cite{ghisello07}).
In this case, care must be used
when determining the optical ``darkness'' of a GRB, since the standard
$\beta_{\rm OX}$ criterium \cite{Jakobsson04} may in this case be easily
violated even for low-redshift, unextinguished afterglows.

\acknowledgments
This work was supported by ASI grant I/R/039/04 and MIUR grant 2005025417. DM
acknowledges support from the Instrument Center for Danish Astrophysics.

\end{document}